\documentclass[12pt]{iopart}
\begin{document}
\title{Primordial and late-time inflation in Brans-Dicke cosmology}
\author{M. Ar\i k and M. C. \c{C}al\i k}
\begin{abstract}
The basic motivation of this work is to attempt to explain the rapid
primordial inflation and the observed slow late-time inflation by using the
Brans-Dicke theory of gravity. We show that the ratio of these two inflation
parameters is proportional to the square root of the Brans-Dicke parameter $%
\omega$ $\left( \omega\gg1\right) $. We also calculate the Hubble parameter $%
H$ and the time variation of the time dependent Newtonian gravitational
constant $G$ for both regimes. The variation of the Hubble parameter
predicted by Brans-Dicke cosmology is shown to be consistent with recent
measurements: The value of $H$ in the late-time future is predicted as 0.86
times the present value of $H$.
\end{abstract}

\address{Bo\~{g}azi\c{c}i Univ., Dept. of Physics, Bebek, Istanbul, Turkey} %
\eads{\mailto{arikm@boun.edu.tr}, \mailto{cem.calik@boun.edu.tr}}

\section{Introduction}

The inflationary universe model whose key feature is a finite period of
primordial rapid exponential expansion has been proposed to resolve a number
of cosmological puzzles, including the horizon, flatness and monopole
problems. In the original or `old inflation model' \cite{guth}, the universe
super-cools into a false vacuum phase and its energy density acts as an
effective cosmological constant which causes an epoch of de-Sitter
(exponential) expansion. In this old inflation model, the de-Sitter
expansion never ends and for a generic first order phase transition, there
appears an energy barrier between the false vacuum and the true vacuum
phases. This problem is known as the`graceful exit' problem. This problem
was avoided with the invention of the new inflationary theory \cite{A.D
Linde}. In this theory, inflation may begin either in the false vacuum, or
in an unstable state at the top of the effective potential. Then the
inflaton field $\phi$ slowly rolls down to the minimum of its effective
potential. The density perturbations produced during the slow-roll inflation
are inversely proportional to $\dot{\phi}$ \cite{mukhanov,Hawking}. Thus the
key difference between the new inflationary scenario and the old one is that
the useful part of inflation in the new scenario, which is responsible for
the homogeneity of our universe, does not occur in the false vacuum state,
where $\dot{\phi}=0$. Although this scenario was so popular in the beginning
of the 80's, it had its own problems. One of them, for example, is that the
inflaton field has an extremely small coupling constant in most versions of
this scenario, so it could not be in thermal equilibrium with other matter
fields. The theory of cosmological phase transitions, which was the basis of
old and new inflation, did not work in this situation. Furthermore,
inflation in this theory begins very late and during the preceding epoch the
universe can easily collapse or become so inhomogeneous that inflation may
never happen \cite{linde4}.

With the invention of the chaotic scenario all problems of old and new
inflation were resolved. According to this scenario, inflation may occur
even in theories with simple potentials such as $V(\phi)\sim\phi^{n}$.
Inflation may begin even if there was no thermal equilibrium in the early
universe, and it may start even at the Planckian density, in which case the
problem of initial conditions for inflation can be easily resolved \cite%
{linde4}. The field in this scenario evolved slowly and at this stage the
energy density of the scalar field, unlike the density of ordinary matter,
remained almost constant, and expansion of the universe continued with a
much greater speed than in the old cosmological theory. Inflation does not
require supercooling and tunnelling from the false vacuum \cite{guth}, or
rolling from an artificially flat top of the effective potential \cite{A.D
Linde}.

All models discussed so far have used field theories at very high energies
to drive inflation. However, inflation may also be generated by changing the
gravitational sector alone ($R^{2}$ inflation) \cite{13- in berkin, 14 in
berkin} or both the gravitational and the matter sectors (Extended
inflation) \cite{16in berkin}. And the well known Jordan-Brans Dicke
theories are widely used in this third type generation of inflation.

Jordan-Brans-Dicke theories are a class of theories in which the effective
gravitational coupling evolves with time, and asymptotically attains a value
of $G$. The strength of the coupling is determined by a scalar field, $\phi$%
, such that asymptotically, it tends to a value $G^{-1}$. The origins of
Brans-Dicke theory is in Mach's principle according to which the property of
inertia of material bodies arises because of their interactions with the
matter distributed in the universe. In the modern context, the Brans-Dicke
theory attempted to rescue the inflationary scenario from some of its
problems.The theory is parameterized by a dimensionless constant $\omega$,
where $\omega\rightarrow\infty$ as Brans-Dicke theory goes over to the
Einstein theory \cite{will}. Present limits of the constant $\omega$ based
on time-delay \cite{reasenberg, C.M.Will, gaztanaga} experiments require $%
\omega>500\gg1$. In the conventional inflationary scenario, the universe
underwent an exponential expansion for a brief period in its early phase.
After the exponential phase is over the universe should transit to the
normal cosmology phase. Within the framework of Einstein-Hilbert action,
there is no satisfactory mechanism by which the universe transits to the
normal phase. It was shown that within the framework of Brans-Dicke gravity,
a constant energy density leads to a rapid power-law expansion instead of
exponential. This is rapid enough to solve the problems in standard
cosmology and at the same time slow enough to make the transition to normal
state possible after the inflationary phase. This has come to be known as
extended inflation \cite{extended, mathiazhan}. Extended inflation
constrains $\omega$ to be less than 25. This bound comes from the fact that
if it is more than 25, there will be much more anisotropy in the Cosmic
Microwave Background Radiation \cite{boomerang} than what is observed today.
This is, however, incompatible with the bound which constrains $\omega$ to
be greater than 500 \cite{reasenberg, gaztanaga}. A large number of
inflationary models were proposed in the framework of multi-scalar tensor
gravity to solve the problem. For instance, the introduction of a potential
for the scalar field $\phi$ and a scalar field dependent coupling constant $%
\omega(\phi)$ solved some problems (\cite{berkin}-\cite{chiba}).

In our work, we start up with a strong link between inflation and
Brans-Dicke \cite{brans} theory of gravity. The proposed model in this work
is simple in that no other phenomenon, such as the domination of the false
vacuum over the scalar field energy density as in the extended inflation
model is used. Since the recent progress in observational cosmology shifted
attention towards experimental verification of various inflationary
theories, the motivation of this work is accelerated with the recent
measurements of the dependence of the Hubble parameter $H=\frac{\dot{a}}{a}$
on the scale size $a(t)$ of the universe as%
\begin{equation}
\left( \frac{H}{H_{0}}\right) ^{2}=\Omega_{\Lambda}+\Omega_{M}\left( \frac{%
a_{0}}{a}\right) ^{3}  \label{parameter}
\end{equation}
\noindent where $\Omega_{\Lambda}\cong0.75$ and $\Omega_{M}\cong0.25$ \cite%
{knop}. In standard cosmology the $\Omega_{\Lambda}$ term would be induced
by a cosmological constant. An immediate question which arises is \ the
physical reason behind this cosmological constant. A universe, expanding
under the sole influence of a cosmological constant $\Lambda=\lambda^{2}$
inflates as $a(t)\sim e^{\lambda t}$. For the present day expansion, $%
\lambda\simeq H_{0}$, whereas for the primordial expansion responsible for
the present large size of the universe, $\lambda$ is much bigger.

In this paper, we will show a natural model where the large ratio of
primordial inflation to present day inflation can be explained by
Brans-Dicke theory which effectively replaces the Newtonian gravitational
constant $G_{N}$ in the Einstein-Hilbert action by a power of the
Brans-Dicke scalar field. The additional kinetic and potential terms of this
scalar field in the action behave effectively as time dependent cosmological
constants. We choose the kinetic term of the scalar field in the Lagrangian
as $\frac{1}{2}g^{\mu\nu }\partial_{\mu}\phi\partial_{\nu}\phi$ in
accordance with ($+---$) metric signature. Then the length dimension of the
scalar field $\phi$ in units where $c=\hbar=1$, is $L^{-1}$ so that $%
G_{N}\sim\phi^{-2}\sim L^{2}$. We make three simple assumptions: (1) The
Brans-Dicke field $\phi$ does not couple to any other field except gravity.
(2) The Lagrangian of the field, in addition to the kinetic term of $\phi$,
contains the simplest chaotic inflation-style potential energy density $%
V(\phi)=\frac{1}{2}m^{2}\phi^{2}$ which is composed only of the scalar field
mass term. (3) $\phi$ evolves with time.

The action is the following;%
\begin{equation}
S=\int d^{4}x\,\sqrt{g}\,\left[ -\frac{1}{8\omega}\,\phi^{2}\,R+\frac{1}{2}%
\,g^{\mu\upsilon}\,\partial_{\mu}\phi\,\partial_{\nu}\phi-V\left(
\phi\right) +L_{M}\right]  \label{action}
\end{equation}
\noindent where $\phi$ represents the Brans-Dicke scalar field and $\omega$
denotes the dimensionless Brans-Dicke parameter taken to be much larger than
$1$ $\left( \omega\gg1\right) $. $L_{M}$, on the other hand, is the matter
lagrangian such that the scalar field $\phi$ does not couple with it. $R$ is
the Ricci \ scalar. For simplicity we also restrict our analysis to the
Robertson Walker metric to emphasize that $\phi$ is necessarily spatially
homogeneous;%
\begin{equation}
ds^{2}=dt^{2}-a^{2}\left( t\right) \,\frac{d\vec{x}^{2}}{\left[ 1+\frac {k}{4%
}\vec{x}^{2}\right] ^{2}}  \label{metric}
\end{equation}
\noindent where $k$\ is the curvature parameter with $k=-1$, $0$, $1$\
corresponding to open, flat, closed universes respectively and $a\left(
t\right) $ is the scale factor of the universe.

After applying the variational procedure to the action and assuming $\phi
=\phi \left( t\right) $ and energy momentum tensor of matter and radiation
excluding $\phi $ is in the perfect fluid form of $T_{\nu }^{\mu
}=diag\left( \rho ,-p,-p,-p\right) $ where $\rho $ is the energy density and
$p$ is the pressure, and also noting that the right hand side of the $\phi $
equation below is set to be zero in accordance with our first assumption on $%
L_{M}$ being independent of $\phi $, the field equations reduce to (dot
denotes $\frac{d}{dt}$)%
\begin{equation}
\frac{3}{4\omega }\,\phi ^{2}\,\left( \frac{\dot{a}^{2}}{a^{2}}+\frac{k}{%
a^{2}}\right) -\frac{1}{2}\,\dot{\phi}^{2}-\frac{1}{2}\,m^{2}\,\phi ^{2}+%
\frac{3}{2\omega }\,\frac{\dot{a}}{a}\,\dot{\phi}\,\phi =\rho  \label{des}
\end{equation}%
\begin{equation}
\frac{-1}{4\omega }\phi ^{2}\left( 2\frac{\ddot{a}}{a}+\frac{\dot{a}^{2}}{%
a^{2}}+\frac{k}{a^{2}}\right) -\frac{1}{\omega }\,\frac{\dot{a}}{a}\,\dot{%
\phi}\,\phi -\frac{1}{2\omega }\,\ddot{\phi}\,\phi -\left( \frac{1}{2}+\frac{%
1}{2\omega }\right) \,\dot{\phi}^{2}+\frac{1}{2}\,m^{2}\,\phi ^{2}=p
\label{pres}
\end{equation}%
\begin{equation}
\ddot{\phi}+3\,\frac{\dot{a}}{a}\,\dot{\phi}+\left[ m^{2}-\frac{3}{2\omega }%
\left( \frac{\ddot{a}}{a}+\frac{\dot{a}^{2}}{a^{2}}+\frac{k}{a^{2}}\right) %
\right] \,\phi =0.  \label{fi}
\end{equation}

\section{Primordial inflation}

In the primordial inflation analysis, we start to solve the field equations (%
\ref{des}-\ref{fi}) for an empty-static universe by setting $\dot{a}=0$ and $%
p=\rho =0$ and get the following vacuum solutions;%
\begin{equation}
\phi =\phi _{o}\,e^{\alpha t}  \label{fay}
\end{equation}%
\begin{equation}
a=a_{\ast }=\mathrm{const}  \label{s2}
\end{equation}%
\begin{equation}
k=1\;(\mathrm{closed\;universe})  \label{k}
\end{equation}%
\noindent where
\begin{equation}
\alpha ^{2}=m^{2}\left( \frac{\omega }{2\omega +3}\right)  \label{alfa}
\end{equation}%
\begin{equation}
a_{\ast }^{2}=\frac{1}{m^{2}}\left( \frac{2\,\omega +3}{3\,\omega +3}\right)
\left( \frac{3}{2\,\omega }\right) .  \label{a star}
\end{equation}%
\noindent From these (\ref{fay}-\ref{k}) solutions, we see that $\phi $
evolves exponentially with expansion parameter $\alpha $ and on the other
hand $a_{\ast }$ is the constant size of this static universe. We regard
that only the closed universe solution is possible as a positive aspect of
this solution since homogeneity of the universe only makes sense if a closed
universe undergoes big-bang. Let us note that, since two variables $\phi
\left( t\right) $ and $a\left( t\right) $ satisfy the three equations (\ref%
{des}-\ref{fi}), these solutions (\ref{fay}, \ref{s2}) are expected to be
stable. To prove this stability we impose the size of the universe $a$ and
the field $\phi $ to be a function of time $t$ as follows;%
\begin{equation}
a=a_{\ast }\left( 1+\varepsilon \,b\left( t\right) \right)  \label{st1}
\end{equation}%
\begin{equation}
\phi =e^{\alpha t}\left( 1+\varepsilon \,\psi \left( t\right) \right)
\label{st2}
\end{equation}%
\noindent where $\varepsilon $ is the perturbation factor ($\varepsilon \ll
1 $) and $b\left( t\right) $, $\psi \left( t\right) $ are perturbation
functions of $a\left( t\right) $ and $\phi \left( t\right) $ respectively.
We get the following differential equations by using (\ref{des}-\ref{fi})
and equalities (\ref{alfa}, \ref{a star})%
\begin{equation}
\dot{\psi}\left( t\right) -\frac{3}{2\,\omega }\,\dot{b}\left( t\right) +%
\frac{3}{2\,\omega \,\alpha \,a_{\ast }^{2}}\,b\left( t\right) =0
\label{dif}
\end{equation}%
\begin{equation}
\ddot{\psi}\left( t\right) +\left( 4+2\omega \right) \,\alpha \,\dot{\psi}%
\left( t\right) +\ddot{b}\left( t\right) +2\,\alpha \,\dot{b}\left( t\right)
-\frac{b\left( t\right) }{a_{\ast }^{2}}=0  \label{dif2}
\end{equation}%
\begin{equation}
\ddot{\psi}\left( t\right) +2\,\alpha \,\dot{\psi}\left( t\right) +3\,\alpha
\,\dot{b}\left( t\right) -\frac{3}{2\,\omega }\,\ddot{b}+\frac{3}{\omega
\,a_{\ast }^{2}}\,\,b=0.  \label{dif3}
\end{equation}%
\noindent Solving (\ref{dif}-\ref{dif3}) simultaneously gives us that $\dot{b%
}=$ $0$ is the only solution which implies $\ddot{b}=0$ and $\dot{\psi}=0$
and $\ddot{\psi}=0$. Namely, this closed universe vacuum solution where $%
a=a_{\ast }=cst$ \ and $\phi \sim e^{\alpha t}$ is stable.

We now investigate how the presence of radiation changes the behavior of the
universe compared to this stable solution. Since solving the equations (\ref%
{des}-\ref{fi}) for the primordial equation of state $p=\frac{1}{3}\rho$ is
hard, we solve (\ref{fi}) for $a\left( t\right) $ by keeping $\phi\sim
e^{\alpha t}$. By changing to the variable $a^{2}\left( t\right)
=\theta\left( t\right) $, (\ref{fi}) turns out to be the following second
order differential equation,%
\begin{equation}
-\frac{3}{4\omega}\,\ddot{\theta}+\frac{3\alpha}{2}\,\dot{\theta}+\left(
\alpha^{2}+m^{2}\right) \,\theta=\frac{3}{2\omega}  \label{dif4}
\end{equation}
\noindent with the following solution for $\theta\left( t\right) $,%
\begin{equation}
\theta\left( t\right) =a^{2}\left( t\right) =\left( \frac{3}{2\omega }%
\right) \left( \frac{1}{\alpha^{2}+m^{2}}\right)
+c_{1\,}e^{\beta_{1}t}+c_{2\,}e^{\beta_{2}t}  \label{soln}
\end{equation}
\noindent where%
\begin{equation}
\beta_{1,2}=\frac{\frac{3\alpha}{2}\,\pm\,\sqrt{\frac{9}{4}\alpha^{2}+\frac {%
3}{\omega}\left( \alpha^{2}+m^{2}\right) }}{\frac{3}{2\omega}}  \label{betas}
\end{equation}
\noindent and $\beta_{1}<0$, $\beta_{2}>0$, $c_{1}$ and $c_{2}$ are
integration constants. Now, if we define the big-bang time as the limit when
$t\rightarrow0$ and $\omega\gg1$, we get $\alpha\simeq\frac{m}{\sqrt{2}}$, $%
\beta_{1}\simeq-2\alpha$, $\beta_{2}\simeq2\omega\alpha$. Equation (\ref%
{soln}) when expanded about $t=0$ becomes%
\begin{equation}
\theta\left( t\right) =a^{2}\left( t\right) =a_{\ast}^{2}\,\left(
1+c_{1}+c_{2}-2\,c_{1\,}\alpha\,t+2\,c_{2\,}\alpha\,\omega\,t\right)
\label{expanded}
\end{equation}
\noindent with the constraint $1+c_{1}+c_{2}=0$ since we want $a^{2}\sim t$
as $t\rightarrow0$. Thus, we end up with the general solution for the scale
size of the universe in the primordial inflation regime with $\omega\gg1$ as;%
\begin{equation}
a^{2}\left( t\right) =a_{\ast}^{2}\left[ 1-\left( 1+c\right) e^{-2\alpha
t}+ce^{2\alpha\omega t}\right] .  \label{general sln}
\end{equation}
\quad

This general solution is important for at least three reasons: (1) It is a
natural solution in the sense that it does not need any special equation of
state for the matter. It is solely deduced from the theory by using the
stable-empty universe solution (\ref{fay}) in the equation (\ref{fi}). (2)
When we examine this inflationary solution concerning as $t\rightarrow0$ and
as $\ t>0$ but not too much, we see that (\ref{general sln}) is both
consistent with $a\left( t\right) \sim\sqrt{t}$ as $t\rightarrow0$ and also
with primordial rapid inflation described by $a\left( t\right) \sim
e^{\alpha\omega t}$ for $\omega $ $\gg1$. (3) We also check for $\omega$ $%
\gg1$ that if we substitute $\phi\sim e^{\alpha t}$ and $a\sim\sqrt{t}$ into
(\ref{des}-\ref{fi}) then the equation of state $p=\frac{1}{3}\rho$ is
satisfied automatically as expected in this regime.

\section{Late-time inflation}

In this section, we analyze how much today's universe is far from late-time
inflation by considering the case of slowly expanding empty universe $%
(\rho=p=0)$ except the $\phi$ field in it. Since the considered universe
should be big, we ignore the curvature parameter $k/a^{2}$ as $a\left(
t\right) $ increases with the expansion of the universe. Under these
considerations, in analogy with the previous section, we put $\ a=e^{\tilde {%
\beta}t}$ and $\phi=e^{\tilde{\alpha}t}$ into (\ref{des}-\ref{fi}) where $%
\tilde{\beta},\tilde{\alpha}$ are new constants to be determined and search
for a stable solution. We get the following coupled equations for $\tilde{%
\beta}$ and $\tilde{\alpha}$;%
\begin{equation}
\tilde{\beta}^{2}-\frac{2}{3}\,\omega\,\tilde{\alpha}^{2}+2\,\tilde{\beta }\,%
\tilde{\alpha}-\frac{2\omega}{3}\,m^{2}=0  \label{bta1}
\end{equation}%
\begin{equation}
\tilde{\beta}^{2}+\left( \frac{2}{3}\omega+\frac{4}{3}\right) \,\tilde {%
\alpha}^{2}+\frac{4}{3}\,\tilde{\beta}\,\tilde{\alpha}-\frac{2\omega}{3}%
\,m^{2}=0  \label{beta}
\end{equation}%
\begin{equation}
\tilde{\beta}^{2}-\frac{\omega}{3}\,\tilde{\alpha}^{2}-\omega\,\tilde{\beta }%
\,\tilde{\alpha}-\frac{\omega}{3}\,m^{2}=0.  \label{beta3}
\end{equation}
\noindent These equations have the solution;%
\begin{equation}
\tilde{\beta}=2\,\left( \omega+1\right) \left( \frac{\omega}{6\omega
^{2}+17\omega+12}\right) ^{1/2}\,m\approx0.8\sqrt{\omega}\,m
\end{equation}%
\begin{equation}
\tilde{\alpha}=\left( \frac{\omega}{6\omega^{2}+17\omega+12}\right)
^{1/2}\,m\approx\frac{0.4}{\sqrt{\omega}}\,m  \label{beta5}
\end{equation}
where the approximations are again for $\omega\gg1$ so that $m\approx\sqrt {2%
}\alpha.$

We see that although the primordial inflation parameter is $0.7\omega m$,
the late-time inflation parameter is found to be $0.8\sqrt{\omega}\,m$.
Namely, a factor $\sqrt{\omega}$ less than the primordial inflation
parameter. This is a very important result in the sense that although there
is an experimental lower bound on $\omega$, there is no upper bound \cite%
{reasenberg} on it, hence in Brans-Dicke cosmology, the late-time inflation
can be as small as one wishes compared to the primordial inflation.

Now, we consider the case where the universe is closed $(k=1)$ and matter
dominated $p\approx 0$. Since solving the field equations (\ref{des}-\ref{fi}%
) for $a\left( t\right) $ and $\phi (t)$ under $p\approx 0$ is hard enough,
we proceed to work by defining the rate of change in $\phi $ as $F\left(
a\right) =\dot{\phi}/\phi $ and Hubble parameter as $H\left( a\right) =\dot{a%
}/a$, and rewriting the right hand side of the field equations(\ref{des}-\ref%
{fi}) in terms of $H$, $F$ and their derivatives with respect to $a$ (prime
denotes $\frac{d}{da}$)%
\begin{equation}
H^{2}-\frac{2\omega }{3}\,F^{2}+2\,H\,F+\frac{1}{a^{2}}-\frac{2\omega }{3}%
\,m^{2}=(\frac{4\omega }{3})\frac{\rho }{\phi ^{2}}  \label{HF1*}
\end{equation}%
\begin{equation}
H^{2}+\left( \frac{2\omega }{3}+\frac{4}{3}\right) \,F^{2}+\frac{4}{3}\,H\,F+%
\frac{2a}{3}\,\left( H\,\acute{H}+H\,\acute{F}\right) +\frac{1}{3a^{2}}-%
\frac{2\omega }{3}\,m^{2}=(\frac{-4\omega }{3})\,\frac{p}{\phi ^{2}}\approx 0
\label{hf2}
\end{equation}%
\begin{equation}
H^{2}-\frac{\omega }{3}\,F^{2}-\omega \,H\,F+a\left( \frac{H\,\acute{H}}{2}-%
\frac{\omega }{3}\,H\acute{F}\right) +\frac{1}{2a^{2}}-\frac{\omega }{3}%
\,m^{2}=0.  \label{HF3}
\end{equation}%
Expanding $F\left( a\right) $ and $H\left( a\right) $ in powers of $\left(
\frac{a_{o}}{a}\right) $ up to third order, where $a_{0}$ is the present
size of a universe,
\begin{equation}
H\left( a\right) =H_{\infty }+H_{2}\left( \frac{a_{0}}{a}\right)
^{2}+H_{3}\left( \frac{a_{0}}{a}\right) ^{3}  \label{H1}
\end{equation}%
\begin{equation}
F\left( a\right) =F_{\infty }+F_{2}\left( \frac{a_{0}}{a}\right)
^{2}+F_{3}\left( \frac{a_{0}}{a}\right) ^{3}  \label{F1}
\end{equation}%
and putting them into (\ref{HF1*}-\ref{HF3}), we get the perturbation
constants defined above for $\left( \omega \gg 1\right) $ :%
\begin{equation}
H_{\infty }=\tilde{\beta}\approx 0.8\sqrt{\omega }\,m  \label{a1}
\end{equation}%
\begin{equation}
H_{2}\approx -\,\frac{1}{2a_{o}^{2}H_{\infty }}\approx -\frac{0.6}{\sqrt{%
\omega }a_{o}^{2}m}\,\approx 0  \label{A2}
\end{equation}%
\begin{equation}
F_{\infty }=\tilde{\alpha}\approx \frac{0.4}{\sqrt{\omega }}\,m  \label{A4}
\end{equation}%
\begin{equation}
F_{2}\approx \frac{3}{4\omega a_{o}^{2}H_{\infty }}\approx \frac{0.9}{\omega
^{3/2}a_{o}^{2}m}\approx 0  \label{A6}
\end{equation}%
\begin{equation}
H_{3}\approx \frac{2\omega }{3}\,F_{3}  \label{A3}
\end{equation}%
\begin{equation}
H_{3}\approx -F_{3}.  \label{(A0)}
\end{equation}%
Up to now, we note that all the constants required in our assumption for $%
H\left( a\right) $ and $F(a)$ in the late-inflation regime are almost
determined from the theory except $H_{3}$ and $F_{3}$. Indeed solving
equations (\ref{A3}) and (\ref{(A0)}) simultaneously gives us $H_{3}$ and $%
F_{3}$ to be zero. But to explain the late-time universe we may assume that $%
\frac{p}{\rho }\ll 1$ rather than $p$ being exactly zero. To overcome this
problem, we use the relation (\ref{A3}) between $H_{3}$ and $F_{3}$ coming
from the equation (\ref{HF3}) which is more exact comparing to the relation (%
\ref{(A0)}) coming from the equation (\ref{hf2}). We also use the classical
Friedman formula which is used for fitting observations of Hubble parameter
to density parameter $\Omega $;%
\begin{equation}
\frac{H^{2}}{H_{0}^{2}}=\Omega _{\Lambda }+\Omega _{R}\left( \frac{a_{0}}{a}%
\right) ^{2}+\Omega _{M}\left( \frac{a_{0}}{a}\right) ^{3}  \label{hubble}
\end{equation}%
\noindent where $\Omega _{\Lambda }$ is the vacuum density parameter, $%
\Omega _{R}$ is the curvature density parameter, $\Omega _{M}$ is the matter
density parameter, and $H_{0}$ is the present Hubble parameter. Using (\ref%
{a1}, \ref{A2}), we rearrange (\ref{H1}) leaving $H_{3}$ as a free parameter
and put into (\ref{hubble}) to get

\begin{equation}
\Omega_{\Lambda}\approx\frac{H_{\infty}^{2}}{H_{0}^{2}}  \label{densp1*}
\end{equation}

\begin{equation}
\Omega_{R}\approx\frac{2H_{\infty}H_{2}}{H_{0}^{2}}  \label{densp2*}
\end{equation}

\begin{equation}
\Omega _{M}\approx \frac{2H_{\infty }H_{3}}{H_{0}^{2}}.  \label{densp3}
\end{equation}%
\noindent\ Using the present observational result \cite{knop} $\Omega
_{M}\approx 0.25$, $\Omega _{\Lambda }\approx 0.75$ and $\Omega _{R}\approx
0 $, we find $H_{3}$ and $F_{3}$ to be,%
\begin{equation}
H_{3}\approx 0.13\sqrt{\omega }m\;\;(\omega \gg 1)  \label{h5}
\end{equation}%
\begin{equation}
F_{3}\approx \,\frac{1.41}{\sqrt{\omega }}m\;\;(\omega \gg 1).  \label{f5}
\end{equation}%
\noindent After finding the perturbation constants explicitly for $H$ and $F$%
, we also find it worthy to determine how the Hubble parameter $H\left(
a\right) =\dot{a}/a$ and the time variation of $G$, where $G$\ is the time
dependent value of the gravitational constant, change\ in the primordial and
late-time regimes compared to their present values. To do so, we use the
fact that since Brans-Dicke gravity becomes identical to Einstein gravity as
$\omega $ approaches infinity, the kinetic term for the scalar field $\frac{1%
}{8\omega }\phi ^{2}$ in the action (\ref{action}) will be the same as that
of the term $1/16\pi G$ in the Hilbert-Einstein action. Using this fact we
get the relation between the scalar field $\phi $ and $G$ as%
\begin{equation}
G^{-1}=\frac{2\pi \phi ^{2}}{\omega }.  \label{G}
\end{equation}%
\noindent Then, putting equations (\ref{a1}), (\ref{A2}), (\ref{h5}) into (%
\ref{H1}) for the present value of the Hubble constant $H_{0}$ and for the
present value of the scale factor of the universe $a_{0}$, gives us the
magnitude of expansion parameter $\alpha $, and mass $m$ of the scalar field
when $\omega \gg $ $1$ as,%
\begin{equation}
\alpha \approx \frac{0.70}{\sqrt{\omega }}H_{0}  \label{ALFA}
\end{equation}%
\begin{equation}
m\approx \frac{1.08}{\sqrt{\omega }}H_{0}.  \label{mass}
\end{equation}

\noindent Similarly, using (\ref{G}) and putting equations (\ref{A4}), (\ref%
{A6}), (\ref{f5}) into (\ref{F1}) for the present value of the scale factor
of the universe $a=a_{0}$, gives us the magnitude of the present value of
the parameter $\dot{G}/G$ when $\omega\gg1$ as,%
\begin{equation}
\left( \frac{\dot{G}}{G}\right) _{0}\approx-\frac{1.81}{\sqrt{\omega}}%
m\approx-\,\frac{1.95}{\sqrt{\omega}}H_{0}.  \label{Gtoday}
\end{equation}
\noindent On the other hand, since $\phi\approx e^{0.7mt}$, $a\approx
e^{0.7m\omega t}$ and $\phi\approx e^{\frac{0.4}{\sqrt{\omega}}\,mt}$, $%
a\approx e^{0.8\sqrt{\omega}\,mt}$ in the primordial and late-time regimes
respectively, using (\ref{G}) we get the parameter $\dot{G}/G$ and the
Hubble parameter $H=\dot{a}/a$ in these regimes as%
\begin{equation}
\left( \frac{\dot{G}}{G}\right) _{primordial}\approx -1.4m\approx-\,\frac{%
1.51}{\sqrt{\omega}}\,H_{0}  \label{gpri}
\end{equation}

\begin{equation}
\left( \frac{\dot{G}}{G}\right) _{late-time}\approx\frac {-0.81}{\sqrt{\omega%
}}m\approx-\,\frac{0.88}{\omega}\,H_{0}  \label{Glate}
\end{equation}

\begin{equation}
\left( H\right) _{primordial}\approx0.7m\omega\approx\allowbreak 0.75\sqrt{%
\omega}\,H_{0}  \label{Hprim}
\end{equation}

\begin{equation}
\left( H\right) _{late-time}\approx0.8\sqrt{\omega}m\approx\allowbreak
0.86H_{0}.  \label{H-late}
\end{equation}

Lastly, we investigate the ratio $\nu =\frac{p}{\rho }$ where $p$ and $\rho $
are pressure and energy density of the late-time universe respectively as $%
\omega \rightarrow \infty $%
\begin{equation}
\nu =\frac{p}{\rho }=\frac{-\frac{\omega +6}{\omega (6\omega +6)}%
\,a_{0}^{-2}-H_{\infty }H_{3}\,\frac{20\omega +21}{\omega (3\omega +3)}\,}{%
\frac{\omega +3}{\omega (2\omega +2)}\,a_{0}^{-2}+2H_{\infty }H_{3}\,}\simeq
0\ \ (\omega \gg 1)\bigskip   \label{p over rho}
\end{equation}%
and find it approaching a value of zero as expected.

\section{Discussion and Conclusion}

In this work we have investigated the nature of the simplest chaotic
inflation-style potential energy density $V(\phi)=\frac{1}{2}m^{2}\phi^{2}$
which is composed only of the scalar field mass term. We have assumed that
and $\phi$ evolves with time in Brans-Dicke Cosmology with a perfect fluid
distribution. We have found a general solution in exponential form for the
size of the universe in primordial regime inflating with an expansion
parameter $\omega\alpha$. We also calculated how the Hubble parameter $%
H\left( a\right) =\dot{a}/a$ and the time variation of $G$, where $G$\ is
the time dependent value of the gravitational constant, change\ in the
primordial and late-time regimes compared to their present values. We note
that the newest measurement \cite{knop} of $\ \Omega_{\Lambda}$ and $%
\Omega_{M}$ has been used as input to derive these results. One interesting
feature is that the predicted present day and primordial values of$\;|\dot
{%
G}/G|$ are comparable whereas the asymptotic value is much smaller. In any
case a measurement of $\dot{G}/G$ will be crucial in determining the
Brans-Dicke parameter $\omega$. On the other hand, the Hubble parameters
predicted by the theory in both regimes have yielded interesting results.
Besides this, the ratio $\nu=\frac{p}{\rho}$ \ is found to be zero as
universe approaches late-time inflation $(\omega\gg1)$.

In the end we can say that the ratio of the primordial and late-time
inflation parameters is proportional to $\sqrt{\omega}$ is the most
appealing feature of Brans-Dicke cosmology. Thus, recent measurements, which
imply that in today's universe$\ \Omega_{\Lambda}\neq0$, require $%
1/\omega\neq0$ and make this model attractive.

\section{Acknowledgments}

We would like to thank the anonymous referee for thoughtful comments and
valuable suggestions on this paper.

\section*{References}

%\begin{harvard}

\end{document}